# Parity-time-symmetric coupled microring lasers operating around an exceptional point

H. Hodaei[1], M. A. Miri[1], A. U. Hassan[1], W. E. Hayenga[1], M. Heinrich[1,2], D. N. Christodouldes[1] and M. Khajavikhan[1*]

[1]CREOL,The College of Optics and Photonics, University of Central Florida, Orlando, Florida, 32816-2700, USA
[2]Institute of Applied Physics, Abbe Center of Photonics, Friedrich-Schiller-Universität Jena, Max-Wien-Platz 1, 07743 Jena, Germany

*Corresponding author: mercedeh@creol.ucf.edu

**The behavior of a parity-time (PT) symmetric coupled microring system is studied when operating in the vicinity of an exceptional point. Using the abrupt phase transition around this point, stable single-mode lasing is demonstrated in spectrally multi-moded micro-ring arrangements.**

The notion of PT-symmetry was first introduced within the context of mathematical physics [1]. It generally implies that even non-Hermitian systems can support entirely real spectra, provided their corresponding Hamiltonians commute with the parity-time (PT) operator. In recent years, several studies have shown that PT symmetric structures can be readily realized in photonic arrangements by exploiting the mathematical isomorphism between the quantum-mechanical Schrödinger equation and the paraxial wave equation of optics [2-7]. In this regard, a necessary albeit not sufficient condition for optical PT symmetry to hold is that the refractive index distribution involved should respect the following relationship $n^*(-\boldsymbol{r}) = n(\boldsymbol{r})$. In other words, the real part of the complex refractive index profile associated with a photonic PT symmetric structure must be an even function of position, while its imaginary component representing gain or loss must have an odd distribution.

By virtue of their non-Hermiticity, PT-symmetric arrangements are capable of supporting exceptional points in their parameter space. Exceptional points, also known as non-Hermitian degeneracies, typically occur in non-conservative physical systems. At these points, the eigenvectors as well as their associated eigenvalues coalesce, resulting in an abrupt phase transition that dramatically changes the behavior of the system [8]. In optical settings, such degeneracies can arise from the interplay of gain and loss in the underlying design. A multitude of recent theoretical and experimental studies has shown different ways of how the presence of exceptional or phase-transition points can be fruitfully utilized to attain new behavior and functionality in non-Hermitian photonic systems, including those respecting PT symmetry [2-17]. Thus far, this methodology has been explored in a number of laser studies where a lower threshold and an inverse pump dependence of the output power have been investigated [18-24]. Lately, the selective breaking of PT symmetry has also been considered for laser mode management applications [23,24]. In particular, in [24], a PT symmetric double ring configuration was proposed as a promising avenue to enforce single longitudinal mode operation in inherently multi-moded semiconductor micro-cavities. This type of system is comprised of two identical ring resonators, where one provides gain while its counterpart is subjected to an appropriate amount of loss.

In this work, our goal is to experimentally characterize the behavior of a parity-time symmetric photonic molecule around the exceptional point. We start with developing a mathematical model for a coupled cavity arrangement showing the emergence of an exceptional point as a result of increasing the gain-loss contrast between the two constituent resonators. We adopt active coupled microring resonators, schematically depicted in Fig. 1, as a platform to perform the related experiments. The coupling strength between the rings is experimentally characterized by examining their emission spectra under uniform pumping. We will then pump the rings differentially and monitor the evolution of their emission spectra as the system transitions through an exceptional point. Finally we show that by properly adjusting the coupling strength as well as gain-loss contrast, an inherently multi-moded microring arrangement can be forced to operate in a single longitudinal resonance. Furthermore, by simply adjusting the ambient temperature, we achieve more than 3 nm continuous wavelength tuning.

Figure 1(a) shows a coupled microring arrangement. In our current analysis, the rings' dimensions (radii, widths and heights) are selected so as to favor the TE polarization while supporting a single transverse mode in the radial direction. The coupling between the two rings is dictated by the mutual overlap of their respective modal fields throughout the interaction region and is set by their separation. The eigenfrequencies associated with the

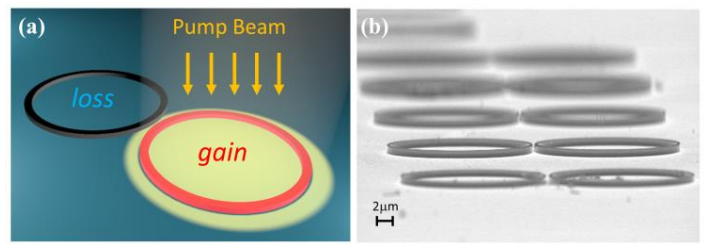

Fig. 1 (a) Schematic of a PT symmetric microring laser. The pump beam is selectively withheld from one of the rings using a knife edge (b) Scanning electron microscope image of a typical set of microring resonator pairs with different separations at an intermediate fabrication step.

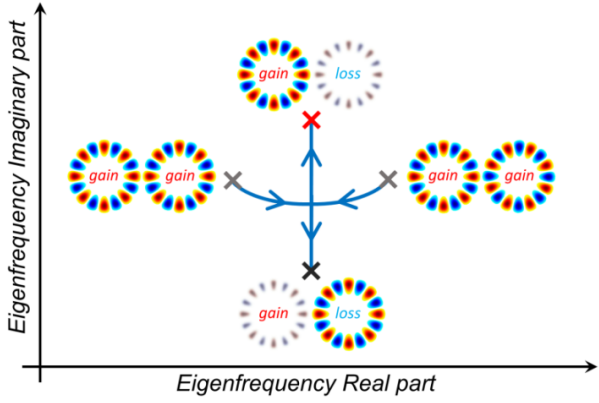

Fig. 2 Evolution of eigenfrequencies as the gain in one of the double active cavities is being gradually replaced with loss. Below the exceptional point, the eigenfrequencies split along the real axis, while keeping the same imaginary components. Above the exceptional point, the real parts of the eigenvalues merge while the imaginary parts bifurcate. At the exceptional point, the eigenfrequencies become completely identical in the complex domain.

supermodes of a double-microring system as depicted in Fig. 1 can be determined through a temporal coupled mode analysis. By considering the n$^{th}$ longitudinal mode in these two rings, one finds that:

$$\omega_n^{(1,2)} = \omega_n + i\left(\frac{\gamma_{a_n} + \gamma_{b_n}}{2}\right) \pm \sqrt{\kappa_n^2 - \left(\frac{\gamma_{a_n} - \gamma_{b_n}}{2}\right)^2} \quad (1)$$

where $\omega_n^{(1,2)}$ represent the eigenfrequencies of the two supermodes of this system, $\omega_n$ is the n$^{th}$ longitudinal resonance frequency of an isolated passive ring, $\gamma_{a_n}$ and $\gamma_{b_n}$ represent the net gain (positive) or loss (negative) in each ring, and $\kappa_n$ stands for the coupling factor.

When both rings are subject to the same level of gain or loss ($\gamma_{a_n} = \gamma_{b_n}$), Eq. 1 reveals that the real parts of the eigenfrequencies split by $2\kappa_n$, while their imaginary components remain degenerate. As a result, uniform pumping scheme can be used to experimentally measure the coupling strength. According to Eq. 1, an interesting situation arises when one of the rings is subject to gain, while the other one to the same amount of loss ($\gamma_{b_n} = -\gamma_{a_n}$). In this case, depending on the relationship between the strength of coupling ($2\kappa_n$) and the gain-loss contrast ($|\gamma_{a_n} - \gamma_{b_n}|$), the splitting either takes place in the real domain or along the imaginary axis. In particular, when the gain-loss contrast between the rings exceeds the coupling ($|\gamma_{a_n} - \gamma_{b_n}| \geq 2\kappa_n$), the real parts of the eigenfrequencies coalesce while the imaginary components lose their degeneracy. The point at which the two eigenfrequencies and their respective eigenvalues fuse ($|\gamma_{a_n} - \gamma_{b_n}| = 2\kappa_n$) is better known as an exceptional point and marks the threshold for parity-time symmetry breaking [2-4].

Figure 2 illustrates the trajectories of the eigenfrequencies associated with the aforementioned coupled microcavity system in the complex domain, as they transition through an exceptional point. In this example, we start with an equally pumped coupled micro-ring arrangement entailing two modes separated by $2\kappa_n$ along the real frequency axis. As one gradually reduces the gain in one of the resonators and replaces it with loss, while the other ring is kept fixed at the initial gain level, the splitting of the resonant frequencies in the real domain monotonically decreases. At the same time, the imaginary components of the eigenfrequencies are reduced due to the lower available net gain. Nevertheless, the imaginary parts of both eigenfrequencies remain equal, indicating that the supermodes are evenly distributed between the two rings. This trend continues until the system reaches the exceptional point. At this juncture, the eigenfrequencies become fully identical in both their real and imaginary parts. Beyond this point and as the gain in the second ring is further reduced, the resonant frequency of the two modes become degenerate, while the imaginary parts bifurcate. These changes reflect the fact that in the broken-symmetry regime one of the supermodes primarily resides in the gain region (therefore experiencing more amplification), while the other one mainly occupies the lossy ring (hence attenuates).

In order to experimentally study the response of a non-Hermitian coupled system around an exceptional point, a number of microring pairs are nano-fabricated following a process similar to that outlined in [25]. The microrings are comprised of six InGaAsP quantum wells. Gain as well as loss is achieved through an appropriate choice of the optical pump intensity supplied to the quantum wells. A scanning electron microscope image of such a microring sample is shown in Fig. 1(b). The rings in our experiments have an outer radius of 10 μm, a width of 500 nm, and a height of 210 nm. The emissions from these structures were visually and spectrally characterized by imaging the surface of the rings onto a CCD camera and by sending the radiation to a spectrometer, respectively. Figure 3 schematically shows the experimental set-up used for characterization of such structures. After passing through a rotating diffuser, the pump beam (1064 nm pulsed laser with a pulse duration of 15 ns and a repetition rate of 290 kHz) is projected to the sample via an NIR objective with a numerical aperture of 0.42. The shaped pump beam can therefore uniformly illuminate a circular area of 40 μm in radius on the sample plane. In order to selectively block certain areas of the sample from pump illumination we utilize a knife-edge as shown in Fig. 3(a). The location of the knife-edge is controlled by a translation stage while the location of its shadow on the sample plane is monitored via a confocal microscope. In this manner, different scenarios of evenly pumped and PT-symmetric coupled microring resonators can be realized as can be seen in Fig. 3 (b).

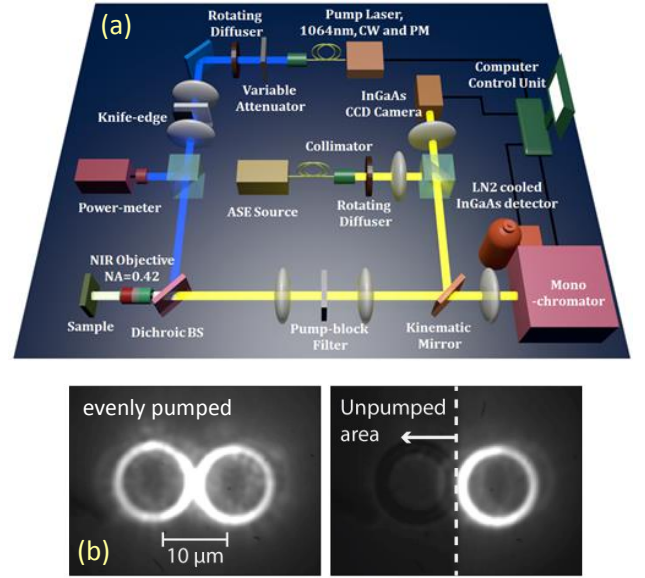

Fig. 3 (a) Micro-photoluminescence characterization set-up equipped with a knife-edge to spatially shape the pump. (b) A knife edge was used to selectively withhold the pump beam from one of the rings.

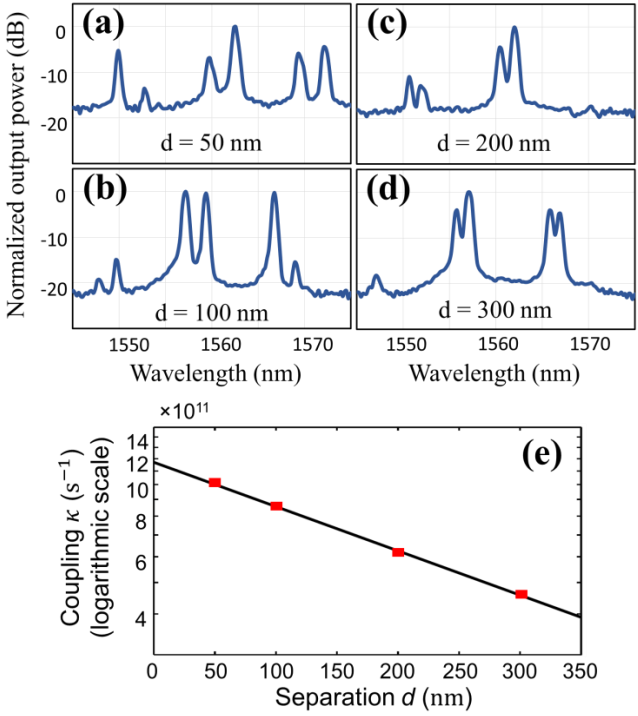

Fig. 4 (a-d) Lasing spectra from evenly pumped microring pairs with separations of 50, 100, 200, and 300 nm, respectively. (e) The coupling coefficient depends exponentially on the separation distance between the two rings. The resolution of the spectrometer is set at 0.4 nm.

As shown in Eq. 1, the exceptional point appears where the coupling strength between the rings becomes equal to the gain-loss contrast ($|\gamma_{a_n} - \gamma_{b_n}| = 2\kappa_n$). Consequently, an accurate measurement of the coupling factor is of fundamental importance for any subsequent investigations. To this end, we exploit the fact that the splitting $\delta\lambda$ of the resonant wavelengths in an evenly pumped pair of rings is directly related to the coupling strength through $\delta\lambda = \kappa\lambda^2/(\pi c)$ [26]. Figures 4(a)-(d) show the lasing spectra for a system of coupled rings separated by 50, 100, 200, and 300 nm, respectively. As expected, the wavelength splitting monotonically decreases as the distance between the two rings is increased. Figure 4(e) confirms the exponential dependence of the coupling factor with respect to the spacing between the rings. For the rings separations at hand, coupling strengths on the order of $\sim 10^{12}$ per second were obtained which is in agreement with simulation results. This level of coupling requires a gain-loss contrast on the order of $\sim 100$ cm$^{-1}$ between the rings to reach the PT symmetry breaking point. In our arrangements, this amount of gain can be readily supplied by the InGaAsP quantum well system with pump powers in the range of a few milliwatts [27]. The variations between the magnitude of the peaks in Figs. 4 (a)-(d) can be attributed to the small detunings due to thermal and/or nonlinear mismatches between the rings [24].

Next, we study the effect of applying a non-uniform pump distribution to observe the transition of the system in the parameter space. For our subsequent experiments, we choose rings (radii: 10 μm, widths: 500 nm, heights: 210 nm) with a separation of 150 nm. Following the scheme illustrated in Fig. 2, we start with an evenly distributed pump (Fig.5 (a)) and gradually introduce loss to one of the rings by partially blocking the pump with a knife edge. Figures 5 (a-d) show the corresponding evolution of the output spectra. As the loss is increased, the two resonant peaks associated with the two supermodes move towards each other, while their respective intensities gradually decrease due to a reduced overall gain. A drastic change, however, happens when the system passes

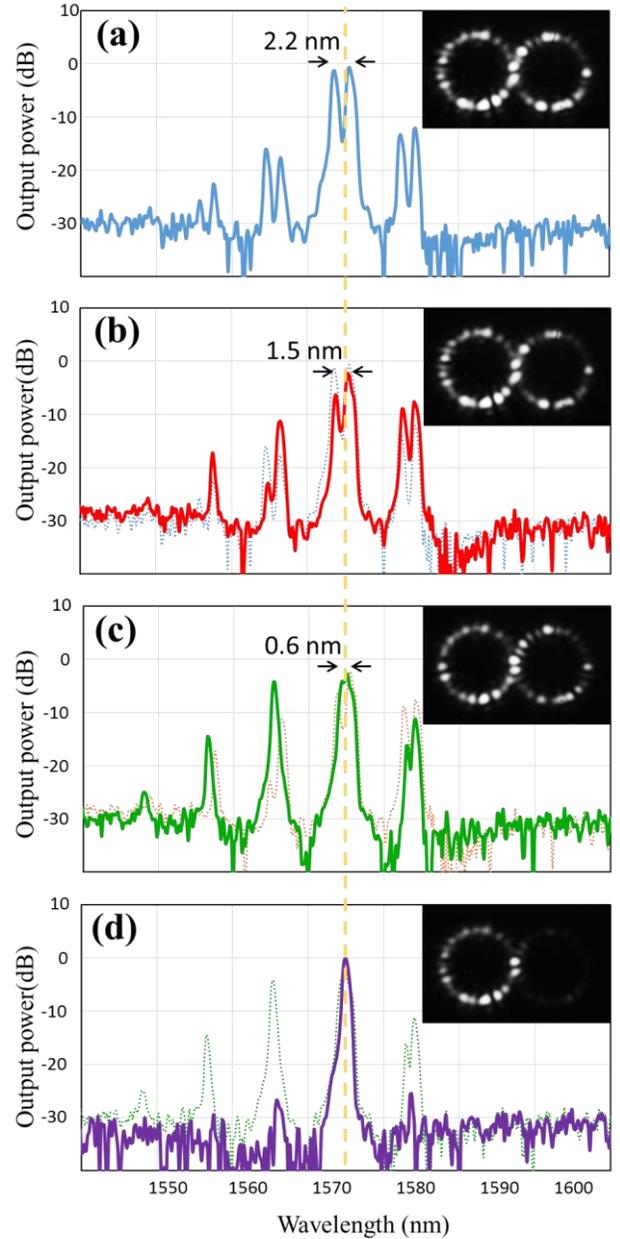

Fig. 5 Evolution of the emission spectra of a double ring structure as a knife-edge gradually covers the right ring, thus introducing loss. The insets show the intensity profile in the two rings collected from the scattered emission. The resolution of the spectrometer is set at 0.4 nm. High-resolution linewidth measurements show that the linewidth in (d) is ~10 GHz.

the exceptional point (Fig. 5(d)). At this point, the two supermodes become spectrally degenerate and only the broken-symmetry mode with the higher Q-factor experiences sufficient amplification to lase. Video files illustrating the continuous evolution of the spectrum as well as the change in the intensity distribution are available as supplementary material. There are two unique characteristics that distinguish this system: i) the PT-broken mode occurs at the center of the two resonant peaks of the evenly pumped system, ii) the PT-mode is not affected by the loss in the un-pumped ring. In [24], the selective breaking of PT-symmetry in double microring systems enabled by the intrinsic lineshape of the gain material has been exploited to enforce single mode lasing in microring arrangements. Similar single-mode lasing behavior can also be observed in Fig. 5(d). As it was shown in [24], single mode lasing can be achieved without detrimental effects on the overall

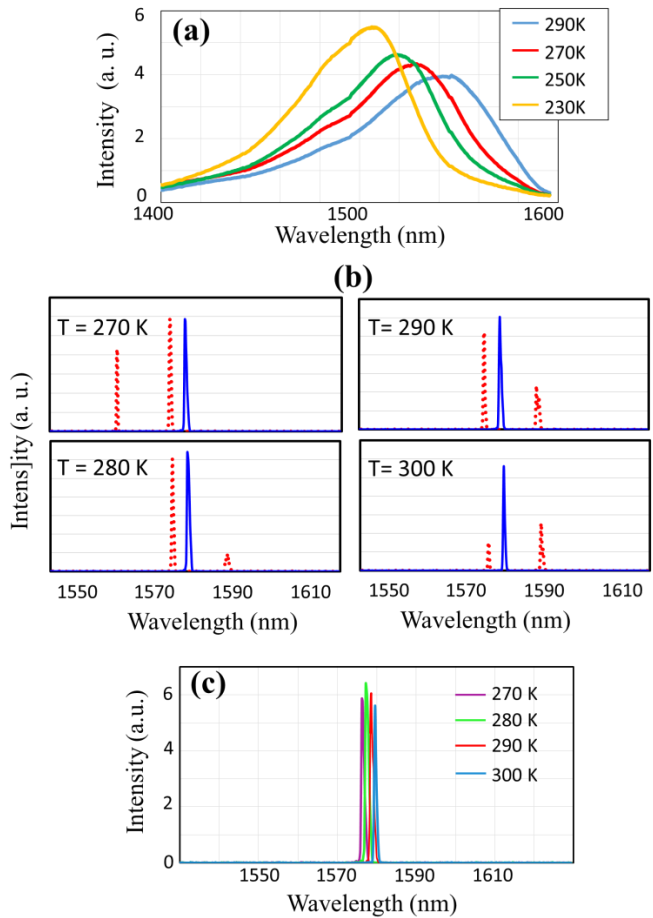

Fig. 6. Continuous wavelength tuning of a PT-symmetric single-mode microring laser. (a) Measured photoluminescence of the InGaAsP QW wafer at different temperatures. (b) Emitted spectra of the PT-symmetric laser (solid blue line) compared to single ring scenario with the same ring dimensions (dotted red line). Continuous single mode tuning is achieved in the PT case while the multi-moded spectrum of a single ring laser undergoes severe mode-hoping by changing the temperature. (c) Emitted spectrum from a PT-symmetric microring structure at temperatures of 270, 280, 290 and 300 K, and a pump power of 0.8, 1, 1.4, and 1.4 mW, respectively. The observed wavelength tuning is approximately 0.1 nmK$^{-1}$. The rings have a radius of 6 µm and a width of 500 nm and are placed 200 nm apart from each other.

lasing efficiency, since the PT-broken mode resides almost exclusively within the active ring.

Finally, we demonstrate wavelength tunability in single mode PT-lasers by utilizing the large intrinsic thermo-optical coefficient of the semiconductor gain material ($dn/dT \sim 10^{-4} K^{-1}$) [28]. In a single ring configuration, the presence of multiple competing modes, often leads to a discontinuous tuning since a slight shift of the amplification envelope may severely impact the ensemble of the modes involved. Figure 6(a) shows the temperature-induced shift of the photoluminescence spectrum of the quantum well active medium. Figure 6(b) compares the evolution of the emission spectra obtained from a conventional ring resonator and the PT-symmetric system as the sample temperature is externally adjusted. Interestingly, the self-adjusting properties of PT-symmetric mode stabilization seem to mitigate mode-hopping effects. Continuous spectral tuning is achieved in a PT arrangement over a range of 3.3 nm by varying the temperature between 270 K and 300 K, while single-modedness is preserved with high fidelity (see Fig. 6(c)). By further increasing the temperature, the single mode resonance jumps by a free spectral range and shifts continuously for another 3.3 nm. This is caused by the shift of the gain spectrum as reported in Fig. 6(a) - resulting in the PT-symmetry breaking to occur at a different longitudinal mode.

In conclusion, we have systematically investigated the lasing dynamics of PT-symmetric microring resonators when operating in the vicinity of an exceptional point. Our work provides insight into the role of non-Hermiticity and the characteristics of phase transitions associated with the spontaneous breaking of PT symmetry. We showed that the ring separation could be used as a design parameter in controlling the performance of a PT-symmetric microring laser. Our study provides an alternative roadmap for designing on-chip light sources which exploit non-Hermiticity as a means to provide stability and enhanced performance in terms of single mode operation and wavelength tunability.

**Funding.** NSF CAREER Award (ECCS-1454531), ARO (W911NF-14-1-0543), NSF (ECCS-1128520), and AFOSR (FA9550-12-1-0148 and FA9550-14-1-0037). M.H. was supported by the German National Academy of Sciences Leopoldina (LPDS 2012-01 and LPDR 2014-03).